\documentclass[prl,twocolumn,showpacs,superscriptaddress,preprintnumbers,amssymb]{revtex4}
\usepackage{graphicx}
\usepackage{dcolumn}
\usepackage{bm}

\newcommand{\beq}{\begin{equation}}
\newcommand{\eeq}{\end{equation}}
\newcommand{\beqn}{\begin{eqnarray}}
\newcommand{\eeqn}{\end{eqnarray}}

\begin{document}

\title{Line defects in Three dimensional Symmetry Protected Topological Phases}

\author{Zhen Bi}

\author{Alex Rasmussen}

\author{Cenke Xu}

\affiliation{Department of physics, University of California,
Santa Barbara, CA 93106, USA}

\begin{abstract}

A 3d symmetry protected topological phase, by definition must have
symmetry protected nontrivial boundary states, namely its 2d
boundary must be either gapless or degenerate. In this work we
demonstrate that once we couple a 3d SPT phase to a lattice
dynamical $Z_2$ gauge field, in many cases the $Z_2$ vison loop
excitation (line defect) can be viewed as a ``1d boundary" of the
3d SPT phase, and this line defect is guaranteed to have gapless
or degenerate spectrum, which is also protected by the symmetry of
the SPT phase.

\end{abstract}

\date{\today}

\maketitle

In the last few years, motivated by the discovery of free fermion
topological insulators protected by time-reversal
symmetry~\cite{kane2005a,kane2005b,bernevig2006,moorebalents2007,fukane,roy2007},
a new class of quantum disordered states, the so called symmetry
protected topological (SPT) phases was
proposed~\cite{wenspt,wenspt2}. Unlike intrinsic topological
phases such as fractional quantum Hall states, a SPT phase is only
nontrivial when the system has certain symmetry $G$. 
A $d-$dimensional SPT phase must have a fully gapped and
nondegenerate spectrum in the bulk, and also a gapless or
degenerate spectrum on its $d-1$ dimensional boundary, when and
only when the Hamiltonian of the system (both in the bulk and the
boundary) has symmetry $G$. In the last two years, SPT phase has
emerged as a new subfield of condensed matter theory, and it has
attracted a lot of attentions and
efforts~\cite{wenspt,wenspt2,levinsenthil,levinstern,liuwen,luashvin,vishwanathsenthil,xu3dspt,xu2dspt,xusenthil,wangsenthil,chenluashvin,maxfisher,yewen1,yewen2}.

Based on the definition of SPT phases, the 2d boundary of a 3d SPT
phase must have nontrivial spectrum. But the properties of a 1d
boundary (or 1d line defect) in a 3d SPT has not been studied yet.
Line defects in ordinary topological insulators have been
discussed before, and it was pointed out that these line defects
do carry gapless modes localized along the
defects~\cite{randefect,randefect2}. In this work we will study
one type of line defects in strongly interacting 3d bosonic SPT
phases, and we will conclude that in many cases, this line defect
in a 3d SPT phase does lead to gapless or degenerate spectrum.

Since so far we do not have explicit lattice model for most of the
SPT phases under study, our work will be based on the effective
field theory of SPT phases. Trivial quantum disordered phases can
be described as the disordered phase of either a Ginzburg-Landau
field theory, or a semiclassical nonlinear sigma model (NLSM)
defined with an order parameter. SPT phases have the same bulk
spectrum and bulk phase diagram as a trivial system, so they can
still be described by NLSMs, and their nontrivial boundary
spectrum can be captured with a topological $\Theta-$term in the
bulk~\cite{vishwanathsenthil,xu3dspt}. It was demonstrated that
the NLSM plus an appropriate topological $\Theta-$term not only
leads to nontrivial boundary physics~\cite{xuludwig}, it also
gives us the correct ground state wave function of the SPT
phase~\cite{xusenthil}. In this work we will focus on several 3d
SPT phases that are described by the same effective field theory,
which is a O(5) Nonlinear Sigma model with a $\Theta-$term at
$\Theta = 2\pi$: \beqn S = \int d^3x d\tau \ \frac{1}{g}
(\partial_\mu \vec{n})^2 + \frac{i \Theta}{ \Omega_4 }
\epsilon_{abcde} n^a \partial_x n^b
\partial_y n^c
\partial_z n^d
\partial_\tau n^e. \label{o5nlsm} \eeqn Here $\vec{n}$ is
a five-component order parameter with unit length. Although these
SPT phases share the same effective field theory, the vector order
parameter $\vec{n}$ transforms differently under symmetry in
different SPT classes.

The order parameter $\vec{n}$ corresponds to certain operators on
the lattice scales, such as spin and boson operators. As long as
the symmetry of the SPT phase contains a $Z_2$ center, $i.e.$ a
$Z_2$ subgroup that commutes with all the other group elements, we
can always modify the Hamiltonian by coupling the lattice
operators to a dynamical $Z_2$ gauge field on the lattice. Since
the matter field $\vec{n}$ is disordered and gapped in the bulk of
these SPT phases, the $Z_2$ gauge field can have a deconfined
phase, which is the phase we will focus on in this paper. The
deconfined phase of a $Z_2$ gauge field introduces line defect in
the system, which is the vison loop of the $Z_2$ gauge field,
$i.e.$ a $\pi-$flux loop in the system. We will argue that in many
SPT phases described by Eq.~\ref{o5nlsm}, the vison loop has a
nontrivial spectrum, namely it is either gapless or degenerate.

\begin{figure}
\includegraphics[width=2.7 in]{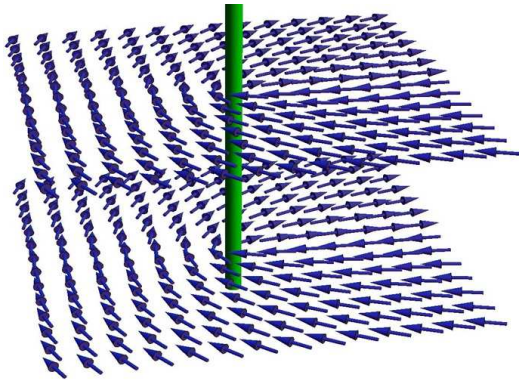}
\caption{A vison loop in example 1 and 2 is bound with a half
vortex line of $b \sim n^1 + in^2$, which leads to a 1+1d O(3)
NLSM with $\Theta  = \pi$ along the vison loop.} \label{defect1}
\end{figure}

{\it Example 1: 3d SPT with $[U(1) \times U(1)]\rtimes Z_2^T$ }

Let us start with a simple example of 3d SPT phase with
$[U(1)\times U(1)]\rtimes Z_2^T$ symmetry, where $Z_2^T$ is the
time-reversal symmetry. This SPT phase is described by
Eq.~\ref{o5nlsm} where $n^1 + in^2 \sim b_1$ and $n^3 + in^4 \sim
b_2$ are two independent boson fields, and $n^5 = \phi$ is an
Ising order parameter that changes sign under
$Z_2^T$~\cite{vishwanathsenthil}: \beqn Z_2^T: b_1 \rightarrow
b_1, \ \ b_2 \rightarrow b_2, \ \ \phi \rightarrow -\phi. \eeqn
Eq.~\ref{o5nlsm} has an enlarged SO(5) symmetry, but we can turn
on extra terms in Eq.~\ref{o5nlsm} which reduce this symmetry down
to physical symmetry $[U(1) \times U(1)] \rtimes Z_2^T$.

We will focus on the SPT phase, namely the phase where the five
component order parameter $\vec{n}$ is completely disordered. We
can couple $b_1$ to a $Z_2$ gauge field, and let us assume this
$Z_2$ gauge field is deep in its deconfined phase, namely the
vison loop excitations of this phase are gapped and dilute.
Although we do not yet have an explicit lattice model for this 3d
SPT, the lattice model of this SPT phase must only contain terms
that are even powers of $b_1$ and $b_1^\dagger$ in order to keep
the U(1) symmetry: $H_0 = \sum_{i,j} - t b^\dagger_{1,i} b_{1,j} +
\cdots$. Thus we can modify this Hamiltonian and couple $b_1$ to a
$Z_2$ gauge field $\sigma^z_{ij}$ defined on the links of the
lattice: $H_g = \sum_{i,j} - t \sigma^z_{ij}b^\dagger_{1,i}
b_{1,j} + \cdots$

Now consider a long vison loop along $z$ axis. This vison loop is
bound with a half-vortex line of $b_1$ (Fig.~\ref{defect1}), and
the vison loop is the core of the half-vortex line.
Along the vison loop (core of half-vortex line), since $n^1$ and
$n^2$ are zero, the effective Lagrangian along the vison loop only
involves a three component unit vector $\vec{n} = (n^3, n^4, n^5)
\sim (\mathrm{Re}[b_2], \mathrm{Im}[b_2], \phi)$. The effective
action along the vison loop reads \beqn \mathcal{S}_v &=& \int dz
d\tau \frac{1}{g^\prime} (\partial_\mu \vec{n})^2 +
\frac{i\Theta_{1d}}{8\pi} \epsilon_{\mu\nu}\epsilon_{abc} n^a
\partial_\mu n^b \partial_\nu n^c, \cr\cr \Theta_{1d} &=& \oint d\vec{l} \
\epsilon_{ef} n^e \partial_l n^f = \pi,  \ \ \ e,f =
1,2.\label{thetap} \eeqn where $l$ is the line coordinate along a
large closed circle around the vison loop.

In Eq.~\ref{thetap}, $\Theta_{1d} = \pi$ is protected by
time-reversal symmetry. Under $Z_2^T$, since a vortex of $b_1$
transforms into an anti-vortex of $b_1$, the derived 1d
$\Theta-$term changes its sign: $\Theta_{1d} \rightarrow -
\Theta_{1d}$, hence Eq.~\ref{thetap} is only time-reversal
invariant at points $\Theta_{1d} = \pi k$ with integer $k$. If we
ignore the physical interpretation of the field $\vec{n}$, this
1+1d NLSM at $\Theta_{1d} = \pi$ (Eq.~\ref{thetap}) can be used to
describe the antiferromagnetic spin-1/2 chain, and based on the
Lieb-Schultz-Mattis (LSM) theorem this 1+1d system must be either
gapless or degenerate~\cite{LSM}. When it is gapless, the vison
loop is described by a 1+1d conformal field theory; degenerate
ground state can be induced by spontaneous time-reversal symmetry
breaking along the vison loop.

Notice that the vison loop is invariant under time-reversal
transformation, because in one plaquette $\pi-$flux and
$-\pi-$flux are equivalent. However, flux lines of other discrete
gauge fields are not necessarily time-reversal invariant, thus if
we couple the same SPT phase to other discrete gauge fields, the
line defects may be fully gapped without degeneracy.

{\it Example 2: 3d SPT with $U(1) \rtimes Z_2$ symmetry}

This SPT phase is also described by Eq.~\ref{o5nlsm}, with the
following transformations of $\vec{n}$: \beqn U(1)&:& b \sim n^1 +
i n^2 \rightarrow e^{i\theta}b, \cr\cr Z_2 &:&, n^1 \rightarrow
n^1, \ \ n^a \rightarrow - n^a, a = 2, \cdots 5. \eeqn The $U(1)
\rtimes
Z_2$ symmetry is a subgroup of SO(5).
Since $b \rightarrow b^\dagger$ under $Z_2$, the U(1) and $Z_2$
symmetries do not commute with each other.

The lattice model of this SPT phase can be constructed using
bosonic rotor operator $b_j \sim \exp(i\phi_j)$ on lattice. The
$Z_2$ symmetry corresponds to the particle-hole transformation of
$b$. $n^3$, $n^4$ and $n^5$ fields in the field theory correspond
to the rotor density operator, which changes sign under $Z_2$
particle-hole transformation. In order to keep the U(1) symmetry,
the lattice Hamiltonian will only involve even powers of $b$, thus
we can couple $b$ to a lattice $Z_2$ gauge field. The rest of the
analysis is very similar to the previous example: the half-vortex
line of $b$ bound with the vison loop will lead to a 1+1d O(3)
NLSM with $\Theta_{1d} = \pi$ along the vison loop, which must be
either gapless or degenerate. $\Theta_{1d} = \pi$ is protected by
the $Z_2$ particle-hole symmetry: under $Z_2$ transformation
$\Theta_{1d} \rightarrow - \Theta_{1d}$, because a vortex of $b$
becomes an anti-vortex under particle-hole transformation.

{\it Example 3: 3d SPT with $Z_2 \times Z_2^T$ symmetry }

In this section we discuss line defects in 3d SPT phases with
discrete symmetries only. Let us take $Z_2 \times Z_2^T$ symmetry
as an example. SPT phases with symmetry $Z_2 \times Z_2^T$ have
$(\mathbb{Z}_2)^3$ classification according to Ref.~\cite{wenspt}.
These eight different phases can be built with three different
basic phases, one is the bosonic topological superconductor with
{\it just} $Z_2^T$ symmetry. The other two correspond to the so
called phase {\bf 1} and {\bf 2} of $U(1) \times Z_2^T$ SPT phases
in Ref.~\cite{vishwanathsenthil}, and by breaking the U(1) down to
its subgroup $Z_2$, the phase {\bf 1} and {\bf 2} in
Ref.~\cite{vishwanathsenthil} become SPT phases with $Z_2 \times
Z_2^T$ symmetry. All these phases are described by the same
effective field theory as Eq.~\ref{o5nlsm}, with a different
transformation of the O(5) vector $\vec{n}$ under the symmetries.

In this section we will take phase {\bf 1} of $Z_2 \times Z_2^T$
SPT phase as an example. In phase {\bf 1} of $Z_2 \times Z_2^T$
SPT phase, the vector $\vec{n}$ transforms as follows: \beqn Z_2
&:& n^a \rightarrow n^a, a = 1 - 3, \ \ \ n^b \rightarrow - n^b, b
= 4, 5. \cr\cr Z_2^T &:& n^a \rightarrow - n^a, a = 1 \cdots 5.
\eeqn
Presumably this SPT phase can be realized in a lattice spin system
with a local Hamiltonian defined with spin operators $(S^x, S^y,
S^z)$ only. The $Z_2$ symmetry can be viewed as the $\pi-$rotation
around $z-$axis. Based on the symmetry transformations, we can
make connection between field theory variables and lattice
operators. For example, in phase {\bf 1} \beqn n^{a}(\vec{x})
&\sim& A_{a} S^x_j + B_a S^y_j + \cdots \ \ a = 4,5; \cr\cr
n^b(\vec{x}) &\sim& C_b S^z_j + D_b (\vec{S}_i \times \vec{S}_j)
\cdot \vec{S}_k  + \cdots \ \ b = 1, 2, 3 \eeqn with real constant
coefficients $A_a$, $B_b$, $C_b$ and $D_b$.

\begin{figure}
\includegraphics[width=3.1 in]{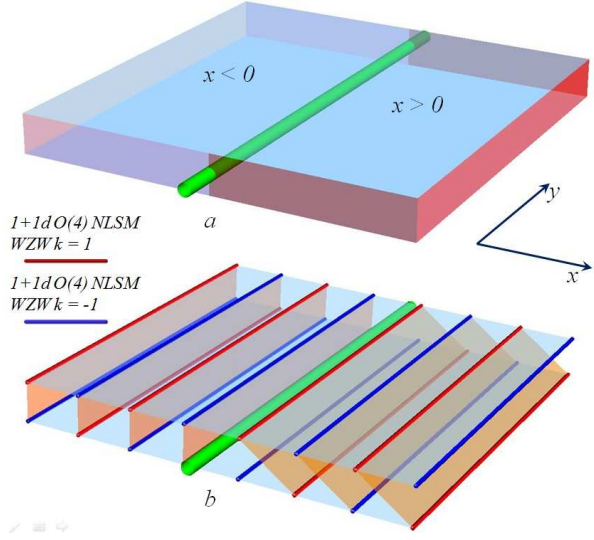}
\caption{A vison line in a 3d SPT phase. $(a)$, the two 2d
boundaries exposed after cutting the SPT phase at $z = 0$. $A(x)$
in Eq.~\ref{coupling} has a domain wall at $x = 0$.
$(b)$, the two boundaries can both be viewed as coupled 1d wires
that are described by 1+1d O(4) NLSM with WZW term at level $k =
\pm 1$ (Eq.~\ref{o4wzw}). } \label{defect2}
\end{figure}

Every term in the lattice Hamiltonian must only have even powers
of $S^x$ and $S^y$ to protect the $Z_2 \times Z_2^T$ symmetry.
Thus we can consistently couple $S^x$ and $S^y$ to a $Z_2$ gauge
theory: $H_g = \sum_{i,j} - t \sigma^z_{ij}S^a_i S^b_j + \cdots $
($a, b = x, y$). With this coupling on the lattice, $n^b$ with $b
= 4,5$ are coupled to the same $Z_2$ gauge field

The simple half-vortex line picture in the previous examples is
not totally applicable here, because there is no U(1) degree of
freedom that can form a vortex around the vison loop. Thus let us
instead consider the following structure: Cut the system open in
the XY plane at $z = 0$, which will expose two boundaries
(Fig.~\ref{defect2}$a$). Both boundaries must have nontrivial
spectrum, and they are both described by a 2+1d NLSM with O(5)
vector $\vec{n}$ plus a Wess-Zumino-Witten (WZW) term at level $k
= \pm 1$ respectively: \beqn && \mathcal{S}_{\alpha } = \int d^2x
d\tau \frac{1}{g} (\partial_\mu \vec{n}_\alpha)^2 \cr\cr &\pm&
\int d^3x \int_0^1 du \frac{2\pi i }{\Omega_4} \epsilon_{abcde}
n^a_{\alpha}
\partial_x n^b_\alpha \partial_y n^c_\alpha \partial_u n^d_\alpha
\partial_\tau n^e_\alpha, \label{wzw}\eeqn Here $\alpha = 1,2$
denotes the top and bottom boundaries exposed. The O(5) WZW term
has level $k = 1$ for top boundary ($\alpha = 1$), and $k = -1$
for bottom boundary ($\alpha = 2$) respectively.
$\vec{n}_\alpha(x, y, \tau, u)$ is an extension of the space-time
configuration $\vec{n}_\alpha (x,y,\tau)$ that satisfies
$\vec{n}_\alpha(x,y, \tau, 0) = (0 ,0 ,0 ,0, 1)$, and
$\vec{n}_\alpha(x,y,\tau, 1) = \vec{n}_\alpha(x,y,\tau)$. The
boundary WZW term can be derived from the bulk $\Theta -$term in
Eq.~\ref{o5nlsm}, because when $\Theta = 2\pi$, the 3+1d bulk
$\Theta-$term can be written as 2+1d WZW terms at boundaries $z =
L$ and $z = 0$: $\Theta-\mathrm{term} = \mathrm{WZW}_{L, k = 1} +
\mathrm{WZW}_{0, k = -1}$.

The symmetry of Eq.~\ref{wzw} needs to be reduced to the physical
symmetry. Let us assume the system energetically favors $n^4$ over
$n^5$, so we can integrate out $n^5_1$ and $n^5_2$ from
Eq.~\ref{wzw} to obtain an effective action for O(4) vectors
$\vec{n}_\alpha = (n^1_\alpha, n^2_\alpha, n^3_\alpha,
n^4_\alpha)$. If the system preserves the $Z_2$ symmetry, then the
expectation values $\langle n^5_1\rangle = \langle n^5_2\rangle =
0$. Now after integrating out $n^5_\alpha$, Eq.~\ref{wzw} is
reduced to two O(4) NLSMs with a $\Theta-$term at $\Theta = \pm
\pi$: \beqn \mathcal{S} &=& \int d^3x \frac{1}{g} (\partial_\mu
\vec{n}_\alpha)^2 \cr\cr &\pm& \int d^3x \frac{i \pi }{12\pi^2}
\epsilon_{abcd}\epsilon_{\mu\nu\rho} n^a_\alpha
\partial_\mu n^b_\alpha \partial_\nu n^c_\alpha \partial_\rho n^d_\alpha. \label{2do4pm}
\eeqn Here $\Theta = \pi$ on the top boundary (or $-\pi$ on the
bottom boundary) is protected by the $Z_2$ symmetry. Detailed
calculation of the $\Theta-$term at the boundary can be found in
Ref.~\cite{vishwanathsenthil,xu3dspt}.

Now let us reglue the two boundaries together, by turning on the
following coupling: \beqn \mathcal{S}_c &=& \int d^2x d\tau \
\sum_{a = 1}^3 B n^a_1(x, \tau) n^a_2(x, \tau) \cr\cr &+& \sum_{b
= 4,5} A(x) n^b_1(x,\tau)n^b_2(x,\tau). \label{coupling}\eeqn The
coupling constant $A(x)$ has a 1d domain wall at $x = z = 0$:
$A(x) < 0 \ \mathrm{for} \ x < 0$, $A(x) > 0 \ \mathrm{for} \
x>0$. For the entire XY plane $B< 0 $. This inter-boundary
coupling corresponds to inserting a vison loop in the XY plane
along the $y-$axis at $x = z = 0$. For the half plane $z = 0, x <
0$, we can identify $\vec{n}_1(x,\tau) = \vec{n}_2(x,\tau) =
\vec{n}(x,\tau)$, and eventually the effective 2d action in this
half-plane is an ordinary O(4) NLSM with no $\Theta-$term. In the
opposite half-plane $z = 0, x > 0$, since $A(x) > 0$, we have
$(n^1_1, n^2_1, n^3_1, n^4_1) = (n^1_2, n^2_2, n^3_2, - n^4_2) =
\vec{n}$, and the effective action for $\vec{n}$ in the half-plane
$x > 0$ is an O(4) NLSM with $\Theta = 2\pi$: \beqn \mathcal{S}_{x
> 0} &=& \int d^3x \frac{1}{g} (\partial_\mu \vec{n})^2 \cr\cr &+&
\int d^3x \frac{i 2\pi }{12\pi^2}
\epsilon_{abcd}\epsilon_{\mu\nu\rho} n^a
\partial_\mu n^b \partial_\nu n^c \partial_\rho n^d. \label{2do4}
\eeqn

Now the vison loop can be viewed as a 1d domain wall of $\Theta$
between $\Theta = 0$ at $x < 0$, and $\Theta = 2\pi$ at $x > 0$.
Although both sides of the domain wall can be driven into a 2d
gapped disordered phase without degeneracy, the domain wall must
have nontrivial spectrum. Using the analysis in
Ref.~\cite{xuludwig}, if both sides of the domain wall are gapped,
this domain wall (vison loop) is described by a 1+1d O(4) NLSM
with a WZW term at level-1: \beqn \mathcal{S} &=& \int dy d\tau \
\frac{1}{g} (\partial_\mu \vec{n})^2 \cr\cr &+& \int d^2x \int_0^1
du\frac{i2\pi}{12\pi^2} n^a
\partial_\mu n^b
\partial_\nu n^c \partial_\rho n^d \epsilon_{abcd}\epsilon_{\mu\nu\rho}.
\eeqn It is well-known that this theory flows to a stable 1+1d
SU(2)$_1$ conformal field theory fixed point under renormalization
group~\cite{witten1984,KnizhnikZamolodchikov1984}, if the system
has a full SO(4) symmetry. In our case, although the symmetry is
much lower than SO(4), no linear term of $n^a$ is allowed by
symmetry. Any bilinear term of $n^a$, even if it is relevant at
the SU(2)$_1$ fixed point, will eventually lead to spontaneous
symmetry breaking and degenerate ground states.


We seek for a more physical picture for the formal calculation
above. Before regluing the boundaries together, the boundaries are
described by O(4) NLSM with $\Theta = \pm \pi$ (Eq.~\ref{2do4pm}).
This theory can be viewed as coupled 1d wires along $y$
direction~\cite{senthilfisher}, and every wire is a 1+1d O(4) NLSM
with a WZW term at level $ k = \pm 1$ (Fig.~\ref{defect2}$b$):
\beqn \mathcal{S}_{\alpha = 1,2, x = j} = \int dy d\tau \
\frac{1}{g} (\partial_\mu \vec{n}_\alpha)^2 \pm \cr\cr  \frac{i
\pi (-1)^j}{12\pi^2} \int dy d\tau \int^1_0 du \
\epsilon_{abcd}\epsilon_{\mu\nu\rho} n^a_\alpha
\partial_\mu n^b_\alpha \partial_\nu n^c_\alpha \partial_\rho n^d_\alpha.
\label{o4wzw}\eeqn If a direct inter-wire coupling $\sum_{\alpha =
1,2} \sum_j \vec{n}(x = j, y, \tau)_\alpha \cdot \vec{n}(x = j+a,
y, \tau)_\alpha $ is turned on ($a$ is the distance between
nearest neighbor wires), each boundary reduces to the 2+1d O(4)
NLSM with $\Theta = \pm \pi$
(Eq.~\ref{2do4pm})~\cite{senthilfisher}.

Now we glue the two boundaries together with a domain wall of
$A(x)$. In the half plane $x < 0$, since $\vec{n}_1 = \vec{n}_2 =
\vec{n}$, two wires on top and bottom boundaries would trivially
gap out due to their coupling with each other (their WZW terms
cancel each other); however, on the other half plane $x > 0$, due
to the opposite sign of inter-boundary coupling, the WZW term of
the top boundary wire $x = j$ will cancel the WZW term of the
bottom boundary wire $x = j+a$. Thus at the domain wall $x = 0$,
there is one 1d wire left which is not gapped by coupling with
other wires.

This picture is very analogous to coupling two spin-1/2 chains
together. Let us consider two spin-1/2 Heisenberg chains along $x$
direction: $H = \sum_{\alpha = 1,2} \sum_{j} \vec{S}_{j,\alpha}
\vec{S}_{j+1,\alpha}$. At $x < 0$ we couple the two chains
antiferromagnetically: $H^\prime = \sum_j J^\prime \vec{S}_{j,1}
\cdot \vec{S}_{j,2}$, while for $x > 0$ we couple the two chains
ferromagnetically: $H^\prime = \sum_j - J^\prime \vec{S}_{j,1}
\cdot \vec{S}_{j,2}$. Then the half-line $x < 0$ can be viewed as
a trivial spin-0 chain, while for $x > 0$ it is the Haldane phase
of a spin-1 chain. Then at the origin $x = 0$ it is guaranteed to
have a dangling spin-1/2 doublet.

The same kind of analysis and conclusion can be applied to the
phase {\bf 2} of SPT phase with $Z_2 \times Z_2^T$ symmetry, where
the O(5) vector $\vec{n}$ transforms as $Z_2 : n^1 \rightarrow
n^1, \ n^b \rightarrow - n^b, b = 2 \cdots 5$; $Z_2^T : n^a
\rightarrow - n^a, a = 1 \cdots 5$. The only difference from phase
{\bf 1} is that, in this case we need to couple $n^b$ with $b = 2
- 5$ to the same $Z_2$ gauge field.

{\it Example 4: Point defect in 2d SPT phase}

Let us now briefly discuss 2d SPT phases. A 2d SPT phase must have
trivial spectrum in the bulk, but gapless or degenerate spectrum
on its 1d boundary. But studies on quantum spin Hall insulator
have suggested that if a point defect is created in a 2d SPT, this
point defect might also change the spectrum. For example, if a
quantum spin Hall insulator is coupled to a $Z_2$ gauge field,
then the vison excitation of this $Z_2$ gauge field must carry a
Kramers doublet~\cite{ranz2,qiz2}.

Here we argue that similar effect also occurs generally in 2d SPT
phases. For instance, let us consider 2d bosonic SPT phase with
$U(1)\rtimes Z_2^T$ symmetry, which is a bosonic version of QSH
insulator. This SPT phase is described by a 2+1d O(4) NLSM with
$\Theta = 2\pi$~\cite{xusenthil} which involves a four component
vector $\vec{n} = (n^1 , n^2, n^3, n^4)$. $n^1 + in^2$ is a boson
rotor that transforms under U(1), and $n^2, n^3, n^4$ all change
sign under $Z_2^T$. Let us couple $n^1$ and $n^2$ to a $Z_2$ gauge
field, and consider a vison at the origin of the 2d system. Then
this vison is bound with a half-vortex of $b$, which leads to a
$0+1d$ O(2) NLSM for $n^3$ and $n^4$ with $\Theta_{0d} = \pi$ at
the origin: $S = \int d\tau \frac{i\pi}{2\pi} \epsilon_{ab} n^a
\partial_\tau n^b $, $a, b = 3,4$. This $0+1d$ model can be solved
exactly, and its ground state is two fold degenerate, which is
precisely a Kramers doublet. This degeneracy is again protected by
time-reversal symmetry. Thus a vison excitation in a $Z_2-$gauged
bosonic quantum spin Hall insulator has the same behavior as the
fermionic QSH state.

If we break the U(1) symmetry down to $Z_2$ (consider 2d SPT phase
with $Z_2 \times Z_2^T$ symmetry), we can still couple $n^1$ and
$n^2$ to a $Z_2$ gauge field. Now a lower dimensional version of
Fig.~\ref{defect2} allows us to study the spectrum of the vison in
the system, and the vison will still be two fold degenerate. The
vison spectrum in this case can also be understood using the
``decorated domain wall" construction of SPT phases discussed in
Ref.~\cite{chenluashvin}. In the 2d $Z_2 \times Z_2^T$ SPT, a
domain wall of the $Z_2$ symmetry is a 1d SPT phase with $Z_2^T$
symmetry, and after coupling the $Z_2$ part to a $Z_2$ gauge
field, a vison is the 0d boundary of the 1d SPT, thus it must be a
2-fold degenerate Kramers doublet. However, none of the defects in
the previous cases discussed in this paper can be analyzed using
the decorated domain wall construction. Our studies based on
effective field theory are more general.

In summary, we study the defects in SPT phases introduced by $Z_2$
gauge field, and in all the cases discussed in this paper the
defect (either line defect in 3d or point defect in 2d) has
nontrivial spectrum. Our study not only reveals a new general
property of SPT phases, it also suggests a possible way of
classifying 3d $Z_2$ topological order enriched by symmetry, based
on the spectrum of its vison line.

CX is supported by the Alfred P. Sloan Foundation, the David and
Lucile Packard Foundation, Hellman Family Foundation, and NSF
Grant No. DMR-1151208. ZB is supported by NSF DMR-1151208.

\bibliography{linedefect}

\end{document}